\documentclass[letterpaper]{article}%
\usepackage{amsmath}
\usepackage{amsfonts}
\usepackage{amssymb}
\usepackage{graphicx}%
\providecommand{\U}[1]{\protect\rule{.1in}{.1in}}
\newtheorem{theorem}{Theorem}

\newtheorem{remark}[theorem]{Remark}

\newcommand{\bpartial}{\mathop{\partial\kern -4pt\raisebox{.8pt}{$|$}}}
\newcommand{\sbpartial}{\tiny\mathop{\partial\kern -4pt\raisebox{.8pt}{$|$}}}
\newcommand{\bra}{\mathopen{[\kern-1.6pt[}}
\newcommand{\ket}{\mathclose{]\kern-1.5pt]}}
\newcommand{\bbra}{\mathopen{[\kern-2.2pt[\kern-2.3pt[}}
\newcommand{\bket}{\mathclose{]\kern-2.1pt]\kern-2.3pt]}}
\newcommand{\sla}{\mbox{\bfseries\slshape a}}
\newcommand{\sle}{\mbox{\bfseries\slshape e}}
\newcommand{\slg}{\mbox{\bfseries\slshape g}}
\newcommand{\sslg}{\mbox{\tiny \bfseries\slshape g}}

\newcommand{\sslm}{\mbox{\tiny \bfseries\slshape m}}

\newcommand{\sitg}{\mbox{\tiny\bfseries\itshape g}}

\newcommand{\slm}{\mbox{\bfseries\slshape m}}

\makeindex
\begin{document}

\title{Nature of the Gravitational Field and its Legitimate Energy-Momentum
Tensor\thanks{This article is based on a talk given by the author at the
9$^{th}$ International Conference on Cliffor Algebras and their Applications
(ICCA9) Weimar, 15-20 July 2011. This version corrects some misprints
appearing in the published version in \emph{Rep. Math. Phys. }\textbf{69},
265-279 (2012).}}
\author{Waldyr A. Rodrigues Jr.\\$\hspace{-0.1cm}$Institute of Mathematics, Statistics and Scientific Computation\\IMECC-UNICAMP\\13083-859 Campinas, SP, Brazil\\e-mail: walrod@ime.unicamp.br\smallskip\ or walrod@mpc.com.br}
\maketitle

\begin{abstract}
In this paper we show how a gravitational field generated by a given
energy-momentum distribution (for all realistic cases) can be represented by
distinct geometrical structures (Lorentzian, teleparallel and non null
nonmetricity spacetimes) or that we even can dispense all those geometrical
structures and simply represent the gravitational field as a field, in the
Faraday's sense, living in Minkowski spacetime. The explicit Lagrangian
density for this theory is given and the field equations (which are\ a set of
four Maxwell's like equations) are shown to be equivalent to Einstein's
equations. We also analyze if the teleparallel formulation can give a
mathematical meaning to \textquotedblleft Einstein's most happy
thought\textquotedblright, i.e. the equivalence principle. Moreover we discuss
the Hamiltonian formalism for for our theory and its relation to one of the
possible concepts for energy of the gravitational field which emerges from it
and the concept of ADM energy. One of the main results of the paper is the
identification in our theory of a legitimate energy-momentum tensor for the
gravitational field expressible through a really nice formula.

\end{abstract}

\section{Introduction}

As well known in General Relativity (GR), a classical field theory of
gravitation, each gravitational field generated by a given energy-momentum
tensor is represented by a \textit{Lorentzian spacetime}, which is a structure
$\langle M,D,%
%TCIMACRO{\TeXButton{g}{\slg}}%
%BeginExpansion
\slg
%EndExpansion
,\tau_{%
%TCIMACRO{\TeXButton{sg}{\sslg}}%
%BeginExpansion
\sslg
%EndExpansion
},\uparrow\rangle$ where $\ M$ is a non compact (locally compact)
$4$-dimensional Hausdorff manifold, $%
%TCIMACRO{\TeXButton{g}{\slg}}%
%BeginExpansion
\slg
%EndExpansion
$ is a Lorentzian metric on $M$ and $D$ is its Levi-Civita connection.
Moreover $M$ is supposed to be oriented by the volume form\ $\tau_{%
%TCIMACRO{\TeXButton{sg}{\sslg}}%
%BeginExpansion
\sslg
%EndExpansion
\ }$ and the symbol $\uparrow$ means that the spacetime is time
orientable\footnote{For details, please consult, e.g., \cite{rodcap2007,sawu}%
.}. From the geometrical objects in the structure $\langle M,D,%
%TCIMACRO{\TeXButton{g}{\slg}}%
%BeginExpansion
\slg
%EndExpansion
,\tau_{%
%TCIMACRO{\TeXButton{sg}{\sslg}}%
%BeginExpansion
\sslg
%EndExpansion
},\uparrow\rangle$ we can calculate the \textit{Riemann curvature tensor}
$\mathbf{R}$ of $D$ and a nontrivial GR model is one in which $\mathbf{R}%
\neq0$. In that way textbooks often say that in GR \textit{spacetime is
curved}. Unfortunately many people mislead the curvature of a connection $D$
on $M$ with the fact that $M$ can eventually be a bent surface in an
(pseudo)Euclidean space with a sufficient number of dimensions\footnote{Any
manifold $M,\dim M=n$ according to the Whitney theorem can be realized as a
submanifold of $\mathbb{R}^{m}$, with $m=2n$. However, if $M$ carries
additional structure the number $m$ in general must be greater than $2n$.
Indeed, it has been shown by Eddington \cite{eddington} that if dim $M=4$ and
if $M$ carries a Lorentzian metric $%
%TCIMACRO{\TeXButton{g}{\slg}}%
%BeginExpansion
\slg
%EndExpansion
$, which moreover satisfies Einstein's equations, then $M$ can be locally
embedded in a (pseudo)euclidean space $\mathbb{R}^{1,9}$. Also, isometric
embedding of a general Lorentzian spacetime would require a lot of extra
dimensions \cite{clarke}. Indeed, a compact Lorentzian manifold can be
embedded isometrically in $\mathbb{R}^{2,46}$ and a non-compact one can be
embedded isometrically in $\mathbb{R}^{2,87}$!}. This confusion leads to all
sort of wishful thinking because many forget that GR does not fix the topology
of $M$ that often must be put \textquotedblleft by hand\textquotedblright%
\ when solving a problem,\ and thus think that they can \textit{bend
}spacetime if they have an appropriate kind of some exotic matter. Worse, the
insistence in supposing that the gravitational field is \textit{geometry} lead
the majority of physicists to relegate the search for the real
\textit{physical nature} of the gravitational field as not important at all
(see a nice discussion of this issue in \cite{laughlin}). What most textbooks
with a few exceptions (see, e.g., the excellent book by Sachs and Wu
\cite{sawu}) forget to say and give a proof to their readers is that in \ the
standard formulation of GR there are \emph{no} genuine conservation laws of
energy-momentum and angular momentum \textit{unless} spacetime has some
\textit{additional} structure which is not present in a general Lorentzian
spacetime \cite{notterod1}. Some textbooks e.g., \cite{mtw} even claim that
energy-momentum conservation for matter plus the gravitational fields is
forbidden due the \textit{equivalence principle}\footnote{We will not discuss
here that most presentations of the equivalence principle are devoid from
mathematical and physical sense. See, e.g., \cite{synge,rodsha1}.} because the
energy-momentum of the gravitational field must be non localizable. Only a few
people tried to develop consistent theories where the gravitational field (at
least from the classical point of view) is simple another field, which like
the electromagnetic field lives in Minkowski spacetime (see a list of
references in \cite{fr}). A field of that nature will be called, in what
follows, a field in Faraday's sense.

Here we want to recall that: \textbf{(i)} the representation of gravitational
fields by Lorentzian spacetimes is not a necessary one, for indeed, there are
some geometrical structures different from $\langle M,D,%
%TCIMACRO{\TeXButton{g}{\slg}}%
%BeginExpansion
\slg
%EndExpansion
,\tau_{%
%TCIMACRO{\TeXButton{sg}{\sslg}}%
%BeginExpansion
\sslg
%EndExpansion
},\uparrow\rangle$ that can equivalently represent such a field; \textbf{(ii)
}The gravitational field can also be nicely represented as a field living in a
fixed background spacetime. The preferred one which seems to describe all
realistic situations is, of course, Minkowski spacetime\footnote{Of course,
the true background spacetime may be eventually a more complicated one, since
that manifold must represent the global topological structure of the universe,
something that is not known at the time of this writing \cite{weeks}. We do
not study this possibility in this paper.} $\langle M\simeq\mathbb{R}%
^{4},\mathring{D},\boldsymbol{\eta},\tau_{\boldsymbol{\eta}},\uparrow\rangle$.

Concerning the possible alternative geometrical models, the particular case
where the connection is \textit{teleparallel} (i.e., it is metric compatible,
has \textit{null} Riemann curvature tensor and \textit{non null }torsion
tensor) will be briefly addressed below (for other possibilities see
\cite{nrr}). What we will show, is that starting with a thoughtful
representation of the gravitational field in terms of \textit{gravitational
potentials} $\mathfrak{g}^{\mathbf{a}}\in\sec$ $%
%TCIMACRO{\tbigwedge \nolimits^{1}}%
%BeginExpansion
{\textstyle\bigwedge\nolimits^{1}}
%EndExpansion
T^{\ast}M\hookrightarrow\sec\mathcal{C}\ell(M,g)$, $\mathbf{a}=0,1,2,3$ and
postulating a convenient Lagrangian density for the gravitational potentials
which \textit{does not use any connection} there is a posteriori different
ways of geometrically representing the gravitational field, such that the
field equations in each representation result equivalent in a precise
mathematical sense to Einstein's field equations. Explicitly we mean by this
statement the following: any \textit{realistic} model of a gravitational field
in GR where that field is represented by\ a \textit{Lorentzian spacetime
}(\textit{with non null Riemann curvature tensor and null torsion tensor
}which is also parallelizable, i.e. admits four \textit{global} linearly
independent vector fields) is \textit{equivalent} to a teleparallel spacetime
(i.e., a spacetime structure equipped with a metrical compatible teleparallel
connection, which has \textit{null }Riemann curvature tensor and \textit{non
null }torsion tensor)\footnote{There are hundreds of papers (as e.g.,
\cite{deandrade}) on the subject, but none (to the best of our knowledge) as
the one presented here.}. The teleparallel possibility follows almost directly
from the results in Section 2 and a recent claim that it can give a
mathematical representation to \textquotedblleft Einstein most happy
though\textquotedblright\ is discussed in Section 3.\medskip

With our teleparallel\ equivalent version of GR and equipped with the powerful
Clifford bundle formalism \cite{fr,rodcap2007} we are able to identify in
Section 4 a legitimate energy momentum tensor for the gravitational field
expressible in a very short and elegant formula.\medskip

Besides this main result we think that another important feature of this paper
is that our representation of the gravitational field by the global $1$-form
fields potentials $\{\mathfrak{g}^{\mathbf{a}}\}$ living on a manifold $M$ and
coupled among themselves and with the matter fields in a specific way (see
below) shows that we can \emph{dispense} with the concept of a connection and
a corresponding geometrical description for that field. The simplest case is
when $M$ is part of Minkowski spacetime structure, in which case the
gravitational field is (like the electromagnetic field) a field in Faraday's
sense\footnote{In \cite{rrr} we even show that when a Lorentzian spacetime
structure $\langle M,D,%
%TCIMACRO{\TeXButton{g}{\slg}}%
%BeginExpansion
\slg
%EndExpansion
,\tau_{%
%TCIMACRO{\TeXButton{sg}{\sslg}}%
%BeginExpansion
\sslg
%EndExpansion
},\uparrow\rangle$ representing a gravitational field in GR possess a Killing
vector field $\mathbf{A}$, then there are Maxwell like equations with well
determined source term satisfied for $F=dA,$ $A=%
%TCIMACRO{\TeXButton{g}{\slg}}%
%BeginExpansion
\slg
%EndExpansion
(\mathbf{A},$ $)$ equivalent to Einstein equation and more, there is a
Navier-Stokes equation equivalent to the Maxwell (like) equations and Einstein
equation.}. In section 5 we present the Hamiltonian formalism for our theory
and discuss the relation of one possible energy concept\footnote{This other
possibility does not define in general a legitimate energy-momentum tensor for
the gravitational field in GR, but defined in our theory in which the
gravitational field is interpreted as a field in the sense of Faraday living
in Minkowski spacetime.} naturally appearing in it and its relation to the
concept of ADM energy. In Section 6 we present the conclusions.

\section{Representation of the Gravitational Field}

Suppose that a $4$-dimensional $M$ manifold is parallelizable, thus admitting
a set of four global linearly independent vector $%
%TCIMACRO{\TeXButton{e}{\sle}}%
%BeginExpansion
\sle
%EndExpansion
_{\mathbf{a}}\in\sec TM$, $\mathbf{a}=0,1,2,3$ fields\footnote{We recall that
$\sec TM$ means section of the tangent bundle and $\sec T^{\ast}M$ means
section of the cotangent bundle. Also $\sec T_{s}^{r}M$ means the bundle of
tensors of type $(r,s)$ and $\sec%
%TCIMACRO{\tbigwedge \nolimits^{r}}%
%BeginExpansion
{\textstyle\bigwedge\nolimits^{r}}
%EndExpansion
T^{\ast}M$ a section of the bundle of $r$-forms fields.} such $\ \{%
%TCIMACRO{\TeXButton{e}{\sle}}%
%BeginExpansion
\sle
%EndExpansion
_{\mathbf{a}}\}$ is a basis for $TM$ and let $\{\mathfrak{g}^{\mathbf{a}%
}\},\mathfrak{g}^{\mathbf{a}}\in\sec T^{\ast}M$ be the corresponding dual
basis ($\mathfrak{g}^{\mathbf{a}}(%
%TCIMACRO{\TeXButton{e}{\sle}}%
%BeginExpansion
\sle
%EndExpansion
_{\mathbf{b}})=\delta_{\mathbf{b}}^{\mathbf{a}}$). Suppose also that not all
the $\mathfrak{g}^{\mathbf{a}}$ are closed, i.e., $d\mathfrak{g}^{\mathbf{a}%
}\neq0$, for a least some $\mathbf{a}=0,1,2,3$. This will be necessary for the
possible interpretations we have in mind for our theory. The $4$-form field
$\mathfrak{g}^{\mathbf{0}}\wedge\mathfrak{g}^{\mathbf{1}}\wedge\mathfrak{g}%
^{\mathbf{2}}\wedge\mathfrak{g}^{\mathbf{3}}$ defines a (positive) orientation
for $M$.

Now, the $\{\mathfrak{g}^{\mathbf{a}}\}$ can be used to define a Lorentzian
metric field in $M$ by defining $%
%TCIMACRO{\TeXButton{g}{\slg}}%
%BeginExpansion
\slg
%EndExpansion
\in\sec T_{2}^{0}M$ by $%
%TCIMACRO{\TeXButton{g}{\slg}}%
%BeginExpansion
\slg
%EndExpansion
:=\eta_{\mathbf{ab}}\mathfrak{g}^{\mathbf{a}}\otimes\mathfrak{g}^{\mathbf{b}}%
$, with the matrix with entries $\eta_{\mathbf{ab}}$ being the diagonal matrix
$(1,-1,-1,-1)$. Then, according to $%
%TCIMACRO{\TeXButton{g}{\slg}}%
%BeginExpansion
\slg
%EndExpansion
$ the $\{\mathbf{e}_{\mathbf{a}}\}$ are orthonormal, i.e., $%
%TCIMACRO{\TeXButton{e}{\sle}}%
%BeginExpansion
\sle
%EndExpansion
_{\mathbf{a}}\underset{%
%TCIMACRO{\TeXButton{sg}{\sslg}}%
%BeginExpansion
\sslg
%EndExpansion
}{\cdot}%
%TCIMACRO{\TeXButton{e}{\sle}}%
%BeginExpansion
\sle
%EndExpansion
_{\mathbf{b}}:=%
%TCIMACRO{\TeXButton{g}{\slg}}%
%BeginExpansion
\slg
%EndExpansion
(%
%TCIMACRO{\TeXButton{e}{\sle}}%
%BeginExpansion
\sle
%EndExpansion
_{\mathbf{a}},%
%TCIMACRO{\TeXButton{e}{\sle}}%
%BeginExpansion
\sle
%EndExpansion
_{\mathbf{b}})=\eta_{\mathbf{ab}}$.

Since the $%
%TCIMACRO{\TeXButton{e}{\sle}}%
%BeginExpansion
\sle
%EndExpansion
_{\mathbf{0}}$ is a \textit{global }time like vector field it follows that it
defines a \textit{time orientation} in $M$ which we denote by $\uparrow$. It
follows that that the 4-tuple $\langle M,%
%TCIMACRO{\TeXButton{g}{\slg}}%
%BeginExpansion
\slg
%EndExpansion
,\tau_{%
%TCIMACRO{\TeXButton{sg}{\sslg}}%
%BeginExpansion
\sslg
%EndExpansion
},\uparrow\rangle$ is part of a structure defining a \textit{Lorentzian
spacetime} and can eventually serve as a \textit{substructure} to model a
gravitational field in GR.

For future use we also introduce $g\in\sec T_{0}^{2}M$ by $g:=\eta
^{\mathbf{ab}}%
%TCIMACRO{\TeXButton{e}{\sle}}%
%BeginExpansion
\sle
%EndExpansion
_{\mathbf{a}}\otimes%
%TCIMACRO{\TeXButton{e}{\sle}}%
%BeginExpansion
\sle
%EndExpansion
_{\mathbf{b}}$, and we write $\mathfrak{g}^{\mathbf{a}}\underset{g}{\cdot
}\mathfrak{g}^{\mathbf{b}}:=g(\mathfrak{g}^{\mathbf{a}},\mathfrak{g}%
^{\mathbf{b}})=\eta^{\mathbf{ab}}$.

Due to the hypothesis that $d\mathfrak{g}^{\mathbf{a}}\neq0$ the commutator of
vector fields $\mathbf{e}_{\mathbf{a}}$, $\mathbf{a}=0,1,2,3$ will in general
satisfy $[%
%TCIMACRO{\TeXButton{e}{\sle}}%
%BeginExpansion
\sle
%EndExpansion
_{\mathbf{a}},%
%TCIMACRO{\TeXButton{e}{\sle}}%
%BeginExpansion
\sle
%EndExpansion
_{\mathbf{b}}]=c_{\mathbf{ab}}^{\mathbf{k}}%
%TCIMACRO{\TeXButton{e}{\sle}}%
%BeginExpansion
\sle
%EndExpansion
_{\mathbf{k}},$where the $c_{\mathbf{ab}}^{\mathbf{k}}$, the structure
coefficients of the basis $\{%
%TCIMACRO{\TeXButton{e}{\sle}}%
%BeginExpansion
\sle
%EndExpansion
_{\mathbf{a}}\}$, and we. easily show that $d\mathfrak{g}^{\mathbf{a}}%
=-\frac{1}{2}c_{\mathbf{kl}}^{\mathbf{a}}\mathfrak{g}^{\mathbf{k}}%
\wedge\mathfrak{g}^{\mathbf{l}}$.

Next, we introduce two different metric compatible connections on $M$, namely
$D$ (the Levi-Civita connection of $%
%TCIMACRO{\TeXButton{g}{\slg}}%
%BeginExpansion
\slg
%EndExpansion
$) and a \textit{teleparallel} connection $\nabla$. Metric compatibility means
that for both connections it is $D%
%TCIMACRO{\TeXButton{g}{\slg}}%
%BeginExpansion
\slg
%EndExpansion
=0$, $\nabla%
%TCIMACRO{\TeXButton{g}{\slg}}%
%BeginExpansion
\slg
%EndExpansion
=0$ \ Now, we put
\begin{align}
D_{%
%TCIMACRO{\TeXButton{e}{\sle}}%
%BeginExpansion
\sle
%EndExpansion
_{\mathbf{a}}}%
%TCIMACRO{\TeXButton{e}{\sle}}%
%BeginExpansion
\sle
%EndExpansion
_{\mathbf{b}}  &  =\omega_{\mathbf{ab}}^{\mathbf{c}}%
%TCIMACRO{\TeXButton{e}{\sle}}%
%BeginExpansion
\sle
%EndExpansion
_{\mathbf{c}},\text{ }D_{%
%TCIMACRO{\TeXButton{e}{\sle}}%
%BeginExpansion
\sle
%EndExpansion
_{\mathbf{a}}}\mathfrak{g}^{\mathbf{b}}=-\omega_{\mathbf{ac}}^{\mathbf{b}%
}\mathfrak{g}^{\mathbf{c}},\nonumber\\
\nabla_{%
%TCIMACRO{\TeXButton{e}{\sle}}%
%BeginExpansion
\sle
%EndExpansion
_{\mathbf{a}}}%
%TCIMACRO{\TeXButton{e}{\sle}}%
%BeginExpansion
\sle
%EndExpansion
_{\mathbf{b}}  &  =0,\text{ \ \ \ \ \ \ \ }\nabla_{%
%TCIMACRO{\TeXButton{e}{\sle}}%
%BeginExpansion
\sle
%EndExpansion
_{\mathbf{a}}}\mathfrak{g}^{\mathbf{b}}=0. \label{7}%
\end{align}

The objects $\omega_{\mathbf{ab}}^{\mathbf{c}}$ are called the
\textit{connection coefficients} of the connection $D$ in the $\{%
%TCIMACRO{\TeXButton{e}{\sle}}%
%BeginExpansion
\sle
%EndExpansion
_{\mathbf{a}}\}$ basis and the objects $\omega_{\mathbf{b}}^{\mathbf{a}}%
\in\sec T^{\ast}M$ defined by $\omega_{\mathbf{b}}^{\mathbf{a}}:=\omega
_{\mathbf{kb}}^{\mathbf{a}}\mathfrak{g}^{\mathbf{k}}$ are called the
\textit{connection} $1$-forms in the $\{%
%TCIMACRO{\TeXButton{e}{\sle}}%
%BeginExpansion
\sle
%EndExpansion
_{\mathbf{a}}\}$ basis. The \textit{connection coefficients} $\varpi
_{\mathbf{ac}}^{\mathbf{b}}$ of $\nabla$ and the connection $1$-forms of
$\nabla$ in the basis $\{%
%TCIMACRO{\TeXButton{e}{\sle}}%
%BeginExpansion
\sle
%EndExpansion
_{\mathbf{a}}\}$ are null according to the second line of Eq.(\ref{7}) and
thus the basis $\{%
%TCIMACRO{\TeXButton{e}{\sle}}%
%BeginExpansion
\sle
%EndExpansion
_{\mathbf{a}}\}$ is called \textit{teleparallel and} the connection $\nabla$
defines an \textit{absolute parallelism} on $M$. Of course, as it is well
known the Riemann curvature tensor of the Levi-Civita connection $D$ of
$\boldsymbol{g}$,\ is in general non null in all points of $M$, but the
torsion tensor of $D$ is zero in all points of $M$. On the other hand the
Riemann curvature tensor of $\nabla$ is null in all points of $M$, whereas the
torsion tensor of $\nabla$ is non null in all points of $M$.

We recall also that for a general connection, say $\mathbf{D}$ on $M$ (not
necessarily metric compatible) the \textit{torsion\/and curvature operations}
and the torsion and \textit{curvature\/}tensors of a given general connection,
say $\mathbf{D}$, are respectively the mappings:%

\begin{align}
\boldsymbol{\rho}  &  :\sec TM\otimes TM\otimes TM\longrightarrow\sec
TM,\nonumber\\
\boldsymbol{\rho}(\boldsymbol{u},\boldsymbol{v},\boldsymbol{w})  &
=D_{\boldsymbol{u}}D_{\boldsymbol{v}}\boldsymbol{w}-D_{\boldsymbol{v}%
}D_{\boldsymbol{u}}\boldsymbol{w}-D_{[\boldsymbol{u,v}}]\boldsymbol{w}%
,\nonumber\\
\mathbf{\tau}  &  \mathbf{:}\sec TM\otimes TM\longrightarrow\sec
TM,\nonumber\\
\mathbf{\tau}(\mathbf{u,v})  &  =\mathbf{D}_{\mathbf{u}}\mathbf{v}%
-\mathbf{D}_{\mathbf{v}}\mathbf{u}-[\mathbf{u,v}]. \label{oprs}%
\end{align}
It is usual to write \cite{choquet} $\boldsymbol{\rho}(\boldsymbol{u}%
,\boldsymbol{v},\boldsymbol{w})=\boldsymbol{\rho}(\boldsymbol{u,v}%
)\boldsymbol{w}$ and $\Theta(\alpha,\mathbf{u,v})=\alpha\left(  \mathbf{\tau
}(u,v)\right)  $ and $\mathbf{R}(\mathbf{w},\alpha,\mathbf{u,v})=\alpha
(\mathbf{\rho(u,v)w})$, for every $\mathbf{u,v,w}\in\sec TM$ and $\alpha
\in\sec\bigwedge^{1}T^{\ast}M$. In particular we write $T_{\mathbf{bc}%
}^{\mathbf{a}}:=\Theta(\mathfrak{g}^{\mathbf{a}},%
%TCIMACRO{\TeXButton{e}{\sle}}%
%BeginExpansion
\sle
%EndExpansion
_{\mathbf{b}}\mathbf{,}%
%TCIMACRO{\TeXButton{e}{\sle}}%
%BeginExpansion
\sle
%EndExpansion
_{\mathbf{c}})$ and $R_{\mathbf{a\;cd}}^{\;\mathbf{b}}:=\mathbf{R}(%
%TCIMACRO{\TeXButton{e}{\sle}}%
%BeginExpansion
\sle
%EndExpansion
_{\mathbf{a}},\mathfrak{g}^{\mathbf{b}},%
%TCIMACRO{\TeXButton{e}{\sle}}%
%BeginExpansion
\sle
%EndExpansion
_{\mathbf{c}},%
%TCIMACRO{\TeXButton{e}{\sle}}%
%BeginExpansion
\sle
%EndExpansion
_{\mathbf{d}})$, and define the Ricci tensor by $Ricci:=R_{\mathbf{ac}%
}\mathfrak{g}^{\mathbf{a}}\otimes\mathfrak{g}^{\mathbf{c}}$ with
$R_{\mathbf{ac}}:=R_{\mathbf{a\;cb}}^{\;\mathbf{b}}=R_{\mathbf{ca}}$.

From now on we imagine $%
%TCIMACRO{\tbigwedge }%
%BeginExpansion
{\textstyle\bigwedge}
%EndExpansion
T^{\ast}M=%
%TCIMACRO{\tbigoplus \nolimits_{r=0}^{4}}%
%BeginExpansion
{\textstyle\bigoplus\nolimits_{r=0}^{4}}
%EndExpansion%
%TCIMACRO{\tbigwedge \nolimits^{r}}%
%BeginExpansion
{\textstyle\bigwedge\nolimits^{r}}
%EndExpansion
T^{\ast}M\hookrightarrow\mathcal{C}\ell(M,g)$, where $\mathcal{C}\ell(M,g)$ is
the Clifford bundle of non homogeneous. differential forms and use the
conventions about the scalar product, left and right contractions,the Hodge
star operator and the Hodge codifferential. operator in $\mathcal{C}\ell(M,g)$
as defined in \cite{rodcap2007,fr}. \medskip

Given that we introduced two different connections $D$ and $\nabla$ defined in
the manifold $M$ we can write \textit{two different pairs} of Cartan's
structure equations. Those pairs describe respectively the geometry of the
structures $\langle M,D,%
%TCIMACRO{\TeXButton{g}{\slg}}%
%BeginExpansion
\slg
%EndExpansion
,\tau_{%
%TCIMACRO{\TeXButton{sg}{\sslg}}%
%BeginExpansion
\sslg
%EndExpansion
},\uparrow\rangle$ and $\langle M,\nabla,%
%TCIMACRO{\TeXButton{g}{\slg}}%
%BeginExpansion
\slg
%EndExpansion
,\tau_{%
%TCIMACRO{\TeXButton{sg}{\sslg}}%
%BeginExpansion
\sslg
%EndExpansion
},\uparrow\rangle$ which will be called respectively a \textit{Lorentzian
spacetime} and a \textit{teleparallel spacetime. }In the case $\langle M,D,%
%TCIMACRO{\TeXButton{g}{\slg}}%
%BeginExpansion
\slg
%EndExpansion
,\tau_{%
%TCIMACRO{\TeXButton{sg}{\sslg}}%
%BeginExpansion
\sslg
%EndExpansion
},\uparrow\rangle$ we write
\begin{equation}
\Theta^{\mathbf{a}}:=d\mathfrak{g}^{a}+\omega_{\mathbf{b}}^{\mathbf{a}}%
\wedge\mathfrak{g}^{\mathbf{b}}=0,\text{ \ \ \ }\mathcal{R}_{\mathbf{b}%
}^{\mathbf{a}}:=d\omega_{\mathbf{b}}^{\mathbf{a}}+\omega_{\mathbf{c}%
}^{\mathbf{a}}\wedge\omega_{\mathbf{b}}^{\mathbf{c}},\nonumber
\end{equation}
where the $\Theta^{\mathbf{a}}\in\sec%
%TCIMACRO{\dbigwedge \nolimits^{2}}%
%BeginExpansion
{\displaystyle\bigwedge\nolimits^{2}}
%EndExpansion
T^{\ast}M\hookrightarrow\sec\mathcal{C}\ell(M,g)$, $\mathbf{a}=0,1,2,3$ and
the $\mathcal{R}_{\mathbf{b}}^{\mathbf{a}}\in\sec%
%TCIMACRO{\dbigwedge \nolimits^{2}}%
%BeginExpansion
{\displaystyle\bigwedge\nolimits^{2}}
%EndExpansion
T^{\ast}M\hookrightarrow\sec\mathcal{C}\ell(M,g)$, $\mathbf{a},\mathbf{b}%
=0,1,2,3$ are respectively the torsion and the curvature $2$-forms of $D$ with%
\begin{equation}
\Theta^{\mathbf{a}}=\frac{1}{2}T_{\mathbf{bc}}^{\mathbf{a}}\mathfrak{g}%
^{\mathbf{b}}\wedge\mathfrak{g}^{\mathbf{c}},\text{ }\mathcal{R}_{\mathbf{b}%
}^{\mathbf{a}}=\frac{1}{2}R_{\mathbf{b\;cd}}^{\;\mathbf{a}}\mathfrak{g}%
^{\mathbf{c}}\wedge\mathfrak{g}^{\mathbf{d}}\text{.} \label{trc}%
\end{equation}

In the case of $\langle M,\nabla,%
%TCIMACRO{\TeXButton{g}{\slg}}%
%BeginExpansion
\slg
%EndExpansion
,\tau_{%
%TCIMACRO{\TeXButton{sg}{\sslg}}%
%BeginExpansion
\sslg
%EndExpansion
},\uparrow\rangle$ since $\varpi_{\mathbf{b}}^{\mathbf{a}}=0$ we have
\begin{equation}
\mathcal{F}^{\mathbf{a}}:=d\mathfrak{g}^{a}+\varpi_{\mathbf{b}}^{\mathbf{a}%
}\wedge\mathfrak{g}^{\mathbf{b}}=d\mathfrak{g}^{a},\text{ \ \ \ }%
\overset{\varpi}{\mathcal{R}_{\mathbf{b}}^{\mathbf{a}}}:=d\varpi_{\mathbf{b}%
}^{\mathbf{a}}+\varpi_{\mathbf{c}}^{\mathbf{a}}\wedge\varpi_{\mathbf{b}%
}^{\mathbf{c}}=0, \label{tele}%
\end{equation}
where the $\mathcal{F}^{\mathbf{a}}\in\sec%
%TCIMACRO{\tbigwedge \nolimits^{2}}%
%BeginExpansion
{\textstyle\bigwedge\nolimits^{2}}
%EndExpansion
T^{\ast}M$, $\mathbf{a}=0,1,2,3$ and the $\overset{\varpi}{\mathcal{R}%
_{\mathbf{b}}^{\mathbf{a}}}$ $\in\sec%
%TCIMACRO{\tbigwedge \nolimits^{2}}%
%BeginExpansion
{\textstyle\bigwedge\nolimits^{2}}
%EndExpansion
T^{\ast}M$, $\mathbf{a},\mathbf{b}=0,1,2,3$ are respectively the torsion and
the curvature $\ 2$-forms of $\nabla$ given by formulas analogous to the ones
in Eq.(\ref{trc}).\medskip

We next postulate that the $\{\mathfrak{g}^{\mathbf{a}}\}$ are the basic
variables representing the gravitation field, and moreover postulate that the
$\{\mathfrak{g}^{a}\}$ interacts with the matter fields through the following
\textit{Lagrangian density}\footnote{We observe that the first term in
Eq.(\ref{11}) can be proved (see, e.g. \cite{rodcap2007}) to be equivalent
just to the Lagrangian density used by Einstein in \cite{einstein4}.}%
\begin{equation}
\mathcal{L=L}_{g}+\mathcal{L}_{m}, \label{10b}%
\end{equation}
where $\mathcal{L}_{m}$ is the matter Lagrangian density and\footnote{A
Lagrangian density equivalent to $\mathcal{L}_{g}$ appearead in \cite{tw}.}%

\begin{equation}
\mathcal{L}_{g}=-\frac{1}{2}d\mathfrak{g}^{\mathbf{a}}\wedge\underset{g}{\star
}d\mathfrak{g}_{\mathbf{a}}+\frac{1}{2}\underset{g}{\delta}\mathfrak{g}%
^{\mathbf{a}}\wedge\underset{g}{\star}\underset{g}{\delta}\mathfrak{g}%
_{\mathbf{a}}+\frac{1}{4}\left(  d\mathfrak{g}^{\mathbf{a}}\wedge
\mathfrak{g}_{\mathbf{a}}\right)  \wedge\underset{g}{\star}\left(
d\mathfrak{g}^{\mathbf{b}}\wedge\mathfrak{g}_{\mathbf{b}}\right)  , \label{11}%
\end{equation}

The form of this Lagrangian is notable, the first term is Yang-Mills like, the
second one is a kind of gauge fixing term and the third term is an
auto-interaction term describing the interaction of the
"vorticities\textquotedblright\ of the potentials (or if you prefer, the
interaction between \textit{Chern-Simons} terms $d\mathfrak{g}^{\mathbf{a}%
}\wedge\mathfrak{g}_{\mathbf{a}}$). Before proceeding we observe that this
Lagrangian is not invariant under arbitrary point dependent Lorentz rotations
of the basic cotetrad fields. In fact, if $\mathfrak{g}^{\mathbf{a}}%
\mapsto\mathfrak{g}^{\prime\mathbf{a}}=\Lambda_{\mathbf{b}}^{\mathbf{a}%
}\mathfrak{g}^{\mathbf{b}}=R\mathfrak{g}^{\mathbf{a}}\tilde{R}$ (where for
each $x\in M$, $\Lambda_{\mathbf{b}}^{\mathbf{a}}(x)\in L_{+}^{\uparrow}$,the
homogeneous and\ orthochronous Lorentz group and $R(x)\in Spin_{1,3}%
\subset\mathbb{R}_{1,3}$)\emph{ }we get that%

\begin{equation}
\mathcal{L}_{g}^{\prime}=-\frac{1}{2}d\mathfrak{g}^{\prime\mathbf{a}}%
\wedge\underset{g}{\star}d\mathfrak{g}_{\mathbf{a}}^{\prime}+\frac{1}%
{2}\underset{g}{\delta}\mathfrak{g}^{\prime\mathbf{a}}\wedge\underset{g}{\star
}\underset{g}{\delta}\mathfrak{g}_{\mathbf{a}}^{\prime}+\frac{1}{4}\left(
d\mathfrak{g}^{\prime\mathbf{a}}\wedge\mathfrak{g}_{\mathbf{a}}^{\prime
}\right)  \wedge\underset{g}{\star}\left(  d\mathfrak{g}^{\prime\mathbf{b}%
}\wedge\mathfrak{g}_{\mathbf{b}}^{\prime}\right)  ,
\end{equation}
differs from $\mathcal{L}_{g}$ by an exact differential. So, the field
equations derived by the variational principle results invariant under a
change of gauge \cite{rodcap2007} and we can always choose a gauge such that
$\underset{g}{\delta}\mathfrak{g}_{\mathbf{a}}=0$.

Now, to derive the field equations directly from Eq.(\ref{11}) using
constrained variations of the $\mathfrak{g}^{\mathbf{a}}$ (i.e., variations
induced by point dependent Lorentz rotations) that do not change the metric
field $\boldsymbol{g}$ is a good exercise in the Clifford calculus, whose
details the interested reader may find\footnote{See errata for reference
\cite{fr} at http://www.ime.unicamp.br/\symbol{126}walrod/plasticwr2012} in
Appendix E of \cite{fr}. The result is:%

\begin{equation}
d\underset{g}{\star}\mathcal{S}_{\mathbf{d}}+\text{ }\underset{g}{\star
}t_{\mathbf{d}}=-\underset{g}{\star}\mathcal{T}_{\mathbf{d}},\label{12}%
\end{equation}
where%
\begin{align}
\underset{g}{\star}t_{\mathbf{d}} &  :=\frac{\partial\mathcal{L}_{g}}%
{\partial\mathfrak{g}^{\mathbf{d}}}=\frac{1}{2}[(\mathfrak{g}_{\mathbf{d}%
}\underset{g}{\lrcorner}d\mathfrak{g}^{\mathbf{a}})\wedge\underset{g}{\star
}d\mathfrak{g}_{\mathbf{a}}-d\mathfrak{g}^{\mathbf{a}}\wedge(\mathfrak{g}%
_{\mathbf{d}}\underset{g}{\lrcorner}\underset{g}{\star}d\mathfrak{g}%
_{\mathbf{a}})]\nonumber\\
&  +\frac{1}{2}d(\mathfrak{g}_{\mathbf{d}}\underset{g}{\lrcorner
}\underset{g}{\star}\mathfrak{g}^{\mathbf{a}})\wedge\underset{g}{\star
}d\underset{g}{\star}\mathfrak{g}_{\mathbf{a}}+\frac{1}{2}(\mathfrak{g}%
_{\mathbf{d}}\underset{g}{\lrcorner}\underset{g}{\star}\mathfrak{g}%
^{\mathbf{a}})\wedge\underset{g}{\star}d\underset{g}{\star}\mathfrak{g}%
_{\mathbf{a}}+\frac{1}{2}d\mathfrak{g}_{\mathbf{d}}\wedge\underset{g}{\star
}\left(  d\mathfrak{g}^{\mathbf{a}}\wedge\mathfrak{g}_{\mathbf{a}}\right)
\nonumber\\
&  -\frac{1}{4}d\mathfrak{g}^{\mathbf{a}}\wedge\mathfrak{g}_{\mathbf{a}}%
\wedge\left[  \mathfrak{g}_{\mathbf{d}}\underset{g}{\lrcorner}%
\underset{g}{\star}\left(  d\mathfrak{g}^{\mathbf{c}}\wedge\mathfrak{g}%
_{\mathbf{c}}\right)  \right]  -\frac{1}{4}\left[  \mathfrak{g}_{\mathbf{d}%
}\underset{g}{\lrcorner}\left(  d\mathfrak{g}^{\mathbf{c}}\wedge
\mathfrak{g}_{\mathbf{c}}\right)  \right]  \wedge\underset{g}{\star}\left(
d\mathfrak{g}^{\mathbf{a}}\wedge\mathfrak{g}_{\mathbf{a}}\right)  ,\label{13}%
\end{align}%
\begin{equation}
\underset{g}{\star}\mathcal{S}_{\mathbf{d}}:=\frac{\partial\mathcal{L}_{g}%
}{\partial d\mathfrak{g}^{\mathbf{d}}}=-\underset{g}{\star}d\mathfrak{g}%
_{\mathbf{d}}-(\mathfrak{g}_{\mathbf{d}}\underset{g}{\lrcorner}%
\underset{g}{\star}\mathfrak{g}^{\mathbf{a}})\wedge\underset{g}{\star
}d\underset{g}{\star}\mathfrak{g}_{\mathbf{a}}+\frac{1}{2}\mathfrak{g}%
_{\mathbf{d}}\wedge\underset{g}{\star}\left(  d\mathfrak{g}^{\mathbf{a}}%
\wedge\mathfrak{g}_{\mathbf{a}}\right)  .\label{14}%
\end{equation}
and the\footnote{We suppose that $\mathcal{L}_{m}$ does not depend explicitly
on the $d\mathfrak{g}^{\mathbf{a}}$.} \
\begin{equation}
\underset{g}{\star}\mathcal{T}_{\mathbf{d}}:=\frac{\partial\mathcal{L}_{m}%
}{\partial\mathfrak{g}^{\mathbf{d}}}=-\underset{g}{\star}T_{\mathbf{d}%
}\label{15}%
\end{equation}
are the energy-momentum $3$-forms of the matter fields\footnote{In reality,
due the conventions used in this paper the true energy-momentum $3$-forms are
$\underset{g}{\star}T_{\mathbf{d}}=-\underset{g}{\star}\mathcal{T}%
_{\mathbf{d}}$.}.

Recalling\ that from Eq.(\ref{tele})\ it is $\mathcal{F}^{\mathbf{a}%
}:=d\mathfrak{g}^{a}$, it is, of course, $d\mathcal{F}^{\mathbf{a}}=0$ and the
field equations (Eq.(\ref{12})) can be written as%
\begin{equation}
d\underset{g}{\star}\mathcal{F}_{\mathbf{d}}=-\underset{g}{\star}%
\mathcal{T}_{\mathbf{d}}-\underset{g}{\star}t_{\mathbf{d}}-\underset{g}{\star
}\mathfrak{h}_{d}, \label{16}%
\end{equation}
where%
\begin{equation}
\mathfrak{h}_{\mathbf{d}}=d\left[  (\mathfrak{g}_{\mathbf{d}}%
\underset{g}{\lrcorner}\underset{g}{\star}\mathfrak{g}^{\mathbf{a}}%
)\wedge\underset{g}{\star}d\underset{g}{\star}\mathfrak{g}_{\mathbf{a}}%
-\frac{1}{2}\mathfrak{g}_{\mathbf{d}}\wedge\underset{g}{\star}\left(
\mathcal{F}^{\mathbf{a}}\wedge\mathfrak{g}_{\mathbf{a}}\right)  \right]  .
\label{17}%
\end{equation}

Recalling the definition of the Hodge coderivative operator acting on sections
of $%
%TCIMACRO{\tbigwedge \nolimits^{r}}%
%BeginExpansion
{\textstyle\bigwedge\nolimits^{r}}
%EndExpansion
T^{\ast}M$ we can write Eq.(\ref{16}) as
\begin{equation}
\underset{g}{\delta}\mathcal{F}^{\mathbf{d}}=-(\mathcal{T}_{\mathbf{\ }%
}^{\mathbf{d}}+\mathbf{t}^{\mathbf{d}}), \label{19}%
\end{equation}
with the $\mathbf{t}^{\mathbf{d}}\in\sec%
%TCIMACRO{\tbigwedge \nolimits^{1}}%
%BeginExpansion
{\textstyle\bigwedge\nolimits^{1}}
%EndExpansion
T^{\ast}M$ given by
\begin{equation}
\mathbf{t}^{\mathbf{d}}:=t^{\mathbf{d}}+\mathfrak{h}^{\mathbf{d}}, \label{20}%
\end{equation}
which are legitimate energy-momentum\footnote{This will become evident in
Section 4 were derive the nice formula for the $\mathbf{t}^{\mathbf{d}}$.}
$1$-form fields for the gravitational field. Note that the \textit{total
energy-momentum} tensor of matter plus the gravitational field is trivially
conserved in our theory, i.e.,
\begin{equation}
\underset{g}{\delta}(\mathcal{T}_{\mathbf{\ }}^{\mathbf{d}}+\mathbf{t}%
^{\mathbf{d}})=0. \label{21}%
\end{equation}

\begin{remark}
Recalling \emph{Eq.(\ref{13})} and \emph{Eq.(\ref{17})} the formula for the
$\mathbf{t}^{\mathbf{d}}$ in \emph{Eq.(\ref{20}) }cannot be, of course, the
nice and short formula we promised to present in the introduction. However, it
is equivalent to the nice formula as shown in Section \emph{4}.
\end{remark}

Recall the similarity of the equations satisfied by the gravitational field to
Maxwell equations. Indeed, in electromagnetic theory on a Lorentzian spacetime
we have only one potential $A\in\sec%
%TCIMACRO{\tbigwedge \nolimits^{1}}%
%BeginExpansion
{\textstyle\bigwedge\nolimits^{1}}
%EndExpansion
T^{\ast}M\hookrightarrow\sec\mathcal{C}\ell(M,g)$ and the field equations are%
\begin{equation}
dF=0,\text{ \ \ }\underset{g}{\delta}F=-J,\text{ } \label{Maxwell}%
\end{equation}
where $F\in\sec%
%TCIMACRO{\tbigwedge \nolimits^{2}}%
%BeginExpansion
{\textstyle\bigwedge\nolimits^{2}}
%EndExpansion
T^{\ast}M\hookrightarrow\sec\mathcal{C}\ell(M,g)$ is the electromagnetic field
and $J\in\sec%
%TCIMACRO{\tbigwedge \nolimits^{1}}%
%BeginExpansion
{\textstyle\bigwedge\nolimits^{1}}
%EndExpansion
T^{\ast}M\hookrightarrow\sec\mathcal{C}\ell(M,g)$ $\ $is the electric current.
As well known the two equations in Eq.(\ref{Maxwell}) can be written (if you
do not mind in introducing the connection $D$ in the game) as a single
equation using the Clifford bundle formalism \cite{rodcap2007}, namely
\begin{equation}
\boldsymbol{\partial}F=J. \label{ma}%
\end{equation}
where we can write $\boldsymbol{\partial}=d-\underset{g}{\delta}%
=\mathfrak{g}^{\mathbf{a}}D_{\mathbf{e}_{\mathbf{a}}}$, where
$\boldsymbol{\partial}$ is the Dirac operator\emph{ }(acting on sections of
$\mathcal{C\ell(}M,g)$).

Now, if you feel uncomfortable in needing four distinct potentials
$\mathfrak{g}^{\mathbf{a}}$ for describing the gravitational field you can put
them together defining a vector valued differential form%
\begin{equation}
\mathfrak{g}=\mathfrak{g}^{\mathbf{a}}\otimes%
%TCIMACRO{\TeXButton{e}{\sle}}%
%BeginExpansion
\sle
%EndExpansion
_{\mathbf{a}}\in\sec%
%TCIMACRO{\dbigwedge \nolimits^{1}}%
%BeginExpansion
{\displaystyle\bigwedge\nolimits^{1}}
%EndExpansion
T^{\ast}M\otimes%
%TCIMACRO{\dbigwedge }%
%BeginExpansion
{\displaystyle\bigwedge}
%EndExpansion
TM\hookrightarrow\sec\mathcal{C}\ell(M,g)\otimes%
%TCIMACRO{\dbigwedge }%
%BeginExpansion
{\displaystyle\bigwedge}
%EndExpansion
TM \label{gpot}%
\end{equation}
and in this case the gravitational field equations are%
\begin{equation}
d\mathcal{F}=0,\text{ \ \ \ }\underset{g}{\delta}\mathcal{F}=-(\mathcal{T}%
_{\mathbf{\ }}+\mathbf{t}), \label{MM1}%
\end{equation}
where $\mathcal{F}=\mathcal{F}^{\mathbf{a}}\otimes%
%TCIMACRO{\TeXButton{e}{\sle}}%
%BeginExpansion
\sle
%EndExpansion
_{\mathbf{a}},\mathcal{T}=\mathcal{T}^{\mathbf{a}}\otimes%
%TCIMACRO{\TeXButton{e}{\sle}}%
%BeginExpansion
\sle
%EndExpansion
_{\mathbf{a}},\mathbf{t=t}^{\mathbf{a}}\otimes%
%TCIMACRO{\TeXButton{e}{\sle}}%
%BeginExpansion
\sle
%EndExpansion
_{\mathbf{a}}$. Again, if you do not mind in introducing the connection $D$ in
the game) by considering the bundle $\mathcal{C\ell(}M,g)\otimes TM$ we can
write the two equations in Eq.(\ref{g2}) as a single equation, i.e.,%
\begin{equation}
\partial\mathcal{F}=\mathcal{T}_{\mathbf{\ }}+\mathbf{t} \label{g3}%
\end{equation}

At this point you may be asking: which is the relation of the theory just
presented with Einstein's GR theory? The answer is that recalling that the
connection $1$-forms $\omega^{\mathbf{cd}}$ of $D$ are given by
\begin{equation}
\omega^{\mathbf{cd}}=\frac{1}{2}\left[  \mathfrak{g}^{\mathbf{d}%
}\underset{g}{\lrcorner}d\mathfrak{g}^{\mathbf{c}}-\mathfrak{g}^{\mathbf{c}%
}\underset{g}{\lrcorner}d\mathfrak{g}^{\mathbf{d}}+\mathfrak{g}^{\mathbf{c}%
}\underset{g}{\lrcorner}(\mathfrak{g}^{\mathbf{d}}\underset{g}{\lrcorner
}d\mathfrak{g}_{\mathbf{a}})\mathfrak{g}^{\mathbf{a}}\right]
\end{equation}
one can show (see \cite{rodcap2007} for details) that the Lagrangian
density\ $\mathcal{L}_{g\text{ }}$ becomes
\begin{equation}
\mathcal{L}_{g\text{ }}=\mathcal{L}_{EH}+d(\mathfrak{g}^{\mathbf{a}}%
\wedge\underset{g}{\star}d\mathfrak{g}_{\mathbf{a}}), \label{26}%
\end{equation}
where
\begin{equation}
\mathcal{L}_{EH}=\frac{1}{2}\mathcal{R}_{\mathbf{cd}}\wedge\underset{g}{\star
}(\mathfrak{g}^{\mathbf{c}}\wedge\mathfrak{g}^{\mathbf{d}}) \label{27}%
\end{equation}
(with $\mathcal{R}_{\mathbf{cd}}$ given by Eq.(\ref{trc})) is the
Einstein-Hilbert Lagrangian density. This permits (with some algebra) to show
that Eqs.(\ref{12}) are indeed equivalent to the usual Einstein equations.

Before ending this section we recall that from Eq.(\ref{12}) we can also
define for our theory a meaningful energy-momentum for the gravitational plus
matter fields \ Indeed, using Stokes theorem for a\ `certain $3$-dimensional
volume', say a ball $B$ we immediately get%

\begin{equation}
P^{\mathbf{a}}:=%
%TCIMACRO{\tint \nolimits_{B}}%
%BeginExpansion
{\textstyle\int\nolimits_{B}}
%EndExpansion
\underset{g}{\star}\left(  \mathcal{T}^{\mathbf{a}}+t^{\mathbf{a}}\right)  =-%
%TCIMACRO{\tint \nolimits_{\partial B}}%
%BeginExpansion
{\textstyle\int\nolimits_{\partial B}}
%EndExpansion
\underset{g}{\star}\mathcal{S}^{\mathbf{a}}. \label{energy}%
\end{equation}

\section{A Comment on Einstein Most Happy Though}

The exercises presented above indicate that a particular geometrical
interpretation for the gravitational field is no more than an option among
many ones. Indeed, it is not necessary to introduce any connection $D$\ or
$\nabla$ on $M$ to have a perfectly well defined theory of the gravitational
field whose field equations are (in a precise mathematical sense) equivalent
to the Einstein field equations. Note that we have not given until now details
on the \textit{global topology} of the world manifold $M$, except that since
we admitted that $M$ carries four global (not all closed) $1$-form fields
$\mathfrak{g}^{\mathbf{a}}$ which defines the object $%
%TCIMACRO{\TeXButton{g}{\slg}}%
%BeginExpansion
\slg
%EndExpansion
$, it follows that $\langle M,D,%
%TCIMACRO{\TeXButton{g}{\slg}}%
%BeginExpansion
\slg
%EndExpansion
,\tau_{%
%TCIMACRO{\TeXButton{sg}{\sslg}}%
%BeginExpansion
\sslg
%EndExpansion
},\uparrow\rangle$ is a \textit{spin manifold} \cite{geroch,rodcap2007}, i.e.,
it admits spinor fields. This, of course, is necessary if the theory is to be
useful in the real world since fundamental matter fields are spinor fields.
The most simple spin manifold is clearly Minkowski spacetime which is
represented by a structure $\langle M=\mathbb{R}^{4\text{ }},\mathring
{D},\mathbf{\eta},\tau_{\eta},\uparrow\rangle$ where $\mathring{D}$ is the
Levi-Civita connection of the Minkowski metric $\mathbf{\eta}$. In that case
it is possible to interpret the gravitational field as a $(1,1)$-extensor
field $\boldsymbol{h}$ which is a field in the Faraday sense living in
$\langle M,\mathring{D},\mathbf{\eta},\tau_{\mathbf{\eta}},\uparrow\rangle$.
The field $\boldsymbol{h}$ is a kind of square of $%
%TCIMACRO{\TeXButton{g}{\slg}}%
%BeginExpansion
\slg
%EndExpansion
$ which has been called in \cite{fr} the plastic distortion field of the
Lorentz vacuum. In that theory the potentials $\mathfrak{g}^{\mathbf{a}%
}=\boldsymbol{h}(\boldsymbol{\gamma}^{\mathbf{a}})$ where $\boldsymbol{\gamma
}^{\mathbf{a}}=\delta_{\mu}^{\mathbf{a}}d\mathtt{x}^{\mu}$, with
$\{\mathtt{x}^{\mu}\}$ global naturally adapted coordinates (in
Einstein-Lorentz-Poincar\'{e} gauge) to the \textit{inertial reference frame}
$\mathbf{I}=\partial/\partial\mathtt{x}^{0}$ according to the structure
$\langle\langle M=\mathbb{R}^{4\text{ }},\mathring{D},\mathbf{\eta},\tau
_{\eta},\uparrow\rangle$, i.e. $\mathring{D}\mathbf{I}=0$. In \cite{fr} we
give the dynamics and coupling of $\boldsymbol{h}$ to the matter fields.

At last we want to comment that, as well known, in Einstein's GR one can
easily distinguish in any \textit{real} \textit{physical laboratory, }i.e.,
not one modelled by a time like worldline (despite some claims on the
contrary) \cite{ohanian} a true gravitational field from an acceleration field
of a given reference frame in Minkowski spacetime. This is because in GR the
\textit{mark} of a real gravitational field is the non null Riemann curvature
tensor of $D$, and the Riemann curvature tensor of the Levi-Civita connection
of $\mathring{D}$ (present in the definition of Minkowski spacetime) is null.
However if we interpret a gravitational field as the torsion $2$-forms on the
structure $(M,\nabla,%
%TCIMACRO{\TeXButton{g}{\slg}}%
%BeginExpansion
\slg
%EndExpansion
,\tau_{%
%TCIMACRO{\TeXButton{sg}{\sslg}}%
%BeginExpansion
\sslg
%EndExpansion
},\uparrow)$ viewed as a deformation of Minkowski spacetime then one can also
interpret an acceleration field of an accelerated reference frame in Minkowski
spacetime as generating an effective teleparallel spacetime
$(M,\overset{e}{\nabla},\eta,\tau_{\eta},\uparrow)$. This can be done as
follows. Let $Z\in\sec TU$, $U\subset M$ with $\mathbf{\eta}(Z,Z)=1$ an
\textit{accelerated reference frame} on Minkowski spacetime. This means (see,
e.g., \cite{rodcap2007} for details) that $%
%TCIMACRO{\TeXButton{a}{\sla}}%
%BeginExpansion
\sla
%EndExpansion
=\mathring{D}_{Z}Z\neq0$. Put $e_{\mathbf{0}}=Z$ and define an accelerated
reference frame as \textit{non} trivial $\ $if $\mathfrak{\vartheta
}^{\mathbf{0}}=\eta(e_{0},)$ is not an exact differential. Next recall that in
$U\subset M$ there always exist \cite{choquet} three other $\mathbf{\eta}%
$-orthonormal vector fields $e_{\mathbf{i}}$, $\mathbf{i}=1,2,3$ such that
$\{e_{a}\}$ is an $\mathbf{\eta}$-orthonormal \ basis for $TU$, i.e.,
$\mathbf{\eta}=\eta_{\mathbf{b}}^{\mathbf{a}}\mathfrak{\vartheta}^{\mathbf{a}%
}\otimes\mathfrak{\vartheta}^{\mathbf{b}}$, where $\{\mathfrak{\vartheta
}^{\mathbf{a}}\}$ is the dual basis\footnote{In general we will also have that
$d\mathfrak{\vartheta}^{\mathbf{i}}\neq0$, $\mathbf{i}=1,2,3$.} of $\{e_{a}%
\}$. We then have, $\mathring{D}_{e_{\mathbf{a}}}e_{\mathbf{b}}=\mathring
{\omega}_{\mathbf{ab}}^{\mathbf{c}}e_{\mathbf{c}},\mathring{D}_{e_{\mathbf{a}%
}}\mathfrak{\vartheta}^{b}=-\mathring{\omega}_{\mathbf{ac}}^{\mathbf{b}%
}\vartheta^{\mathbf{c}}.$

What remains in order to be possible to interpret an acceleration field as a
kind of\ `gravitational field' is to introduce on $M$ a $\eta$-metric
compatible connection $\overset{e}{\nabla}$ such that the $\{e_{a}\}$ is
teleparallel according to it, i.e., $\overset{e}{\nabla}_{e_{\mathbf{a}}%
}e_{\mathbf{b}}=0,\overset{e}{\nabla}_{e_{\mathbf{a}}}\mathfrak{\vartheta
}^{\mathbf{b}}=0$. Indeed, with this connection the structure $\langle
M\simeq\mathbb{R}^{4},\overset{e}{\nabla},\mathbf{\eta},\tau_{\eta}%
,\uparrow\rangle$ has null Riemann curvature tensor but a non null torsion
tensor, whose components are related with the components of the acceleration $%
%TCIMACRO{\TeXButton{a}{\sla}}%
%BeginExpansion
\sla
%EndExpansion
$ and with the other coefficients $\mathring{\omega}_{\mathbf{ab}}%
^{\mathbf{c}}$ of the connection $\mathring{D}$, which describe the motion on
Minkowski spacetime of a \textit{grid} represented by the orthonormal frame
$\{e_{a}\}$. Sch\"{u}cking \cite{schu} \ thinks that such a description of the
gravitational field makes Einstein most happy though, i.e., the equivalence
principle (understood as equivalence between acceleration and gravitational
field) a legitimate mathematical idea. However, a \textit{true} gravitational
field must satisfy (at least with good approximation) Eq.(\ref{16}), whereas
there is no single reason for an acceleration field to satisfy that equation.

\section{The Nice Formula for the Legitimate Energy-Momentum Tensor of the
Gravitational Field}

Taking into account that $\mathcal{F}^{\mathbf{d}}=d\mathfrak{g}%
^{d}=\boldsymbol{\partial\wedge}\mathfrak{g}^{d}$ we return to Eq.(\ref{19})
and write it as%
\begin{equation}
\boldsymbol{\partial}^{2}\mathfrak{g}^{d}=\mathcal{T}^{\mathbf{d}%
}+\mathfrak{t}^{\mathbf{d}},\label{n1}%
\end{equation}
with $\mathfrak{t}^{\mathbf{d}}=\mathbf{t}^{\mathbf{d}}-d\delta\mathfrak{g}%
^{\mathbf{d}}$. Next we recall that in the Clifford bundle formalism the
operator $\boldsymbol{\partial}^{2}$ (the Hodge D'Alembertian) has two non
equivalent decompositions, namely, for each $M\in\sec%
%TCIMACRO{\tbigwedge }%
%BeginExpansion
{\textstyle\bigwedge}
%EndExpansion
T^{\ast}M\hookrightarrow\sec\mathcal{C}\ell(M,g)$ we have%
\begin{align}
\boldsymbol{\partial}^{2}M &  =-(d\underset{g}{\delta}+\underset{g}{\delta
}d)M\nonumber\\
&  =\boldsymbol{\partial}\wedge\boldsymbol{\partial}M+\boldsymbol{\partial
}\cdot\boldsymbol{\partial}M\label{n2}%
\end{align}
where $\boldsymbol{\partial}\wedge\boldsymbol{\partial}$ is an extensorial
operator called the Ricci operator and $\boldsymbol{\partial}\cdot
\boldsymbol{\partial}$ is the covariant D'Alembertian operator. We have
\begin{equation}
\boldsymbol{\partial}\wedge\boldsymbol{\partial}\mathfrak{g}^{\mathbf{d}%
}=\mathcal{R}^{\mathbf{d}},\label{n3}%
\end{equation}
where the $\mathcal{R}^{\mathbf{d}}=R_{\mathbf{a}}^{\mathbf{d}}\mathfrak{g}%
^{\mathbf{d}}\in\sec%
%TCIMACRO{\tbigwedge ^{1}}%
%BeginExpansion
{\textstyle\bigwedge^{1}}
%EndExpansion
T^{\ast}M\hookrightarrow\sec\mathcal{C}\ell(M,g)$ (with $R_{\mathbf{a}%
}^{\mathbf{d}}$ the components of the Ricci tensor) are called the Ricci
$1$-form fields. Then we can write Eq.(\ref{n1}) as%
\begin{equation}
\boldsymbol{\partial}\wedge\boldsymbol{\partial}\mathfrak{g}^{\mathbf{d}%
}+\boldsymbol{\partial}\cdot\boldsymbol{\partial}\mathfrak{g}^{\mathbf{d}%
}=\mathcal{T}^{\mathbf{d}}+\mathfrak{t}^{\mathbf{d}},\label{n4}%
\end{equation}
or%
\begin{equation}
\mathcal{R}^{\mathbf{d}}+\boldsymbol{\partial}\cdot\boldsymbol{\partial
}\mathfrak{g}^{\mathbf{d}}=\mathcal{T}^{\mathbf{d}}+\mathfrak{t}^{\mathbf{d}%
}.\label{n5}%
\end{equation}
Now, we recall that Einstein equation in components form is%
\begin{equation}
R_{\mathbf{d}}^{\mathbf{a}}-\frac{1}{2}\delta_{\mathbf{d}}^{\mathbf{a}%
}R=-\boldsymbol{T}_{\mathbf{d}}^{\mathbf{a}}=\mathcal{T}_{\mathbf{d}%
}^{\mathbf{a}}\label{n6}%
\end{equation}
from where it follows immediately that%
\begin{equation}
\mathcal{R}^{\mathbf{d}}-\frac{1}{2}R\mathfrak{g}^{\mathbf{d}}=\mathcal{T}%
^{\mathbf{d}}.\label{n7}%
\end{equation}
Then
\begin{equation}
\mathcal{R}^{\mathbf{d}}+\boldsymbol{\partial}\cdot\boldsymbol{\partial
}\mathfrak{g}^{\mathbf{d}}=\mathcal{T}^{\mathbf{d}}+\frac{1}{2}R\mathfrak{g}%
^{\mathbf{d}}+\boldsymbol{\partial}\cdot\boldsymbol{\partial}\mathfrak{g}%
^{\mathbf{d}},\label{n8}%
\end{equation}
and comparing Eq.(\ref{n4}) with Eq.(\ref{n8}) we get
\begin{equation}
\mathfrak{t}^{\mathbf{d}}=\frac{1}{2}R\mathfrak{g}^{\mathbf{d}}%
+\boldsymbol{\partial}\cdot\boldsymbol{\partial}\mathfrak{g}^{\mathbf{d}%
},\label{n9}%
\end{equation}
and%
\begin{equation}
\mathbf{t}^{\mathbf{d}}=\frac{1}{2}R\mathfrak{g}^{\mathbf{d}}%
+\boldsymbol{\partial}\cdot\boldsymbol{\partial}\mathfrak{g}^{\mathbf{d}%
}+d\delta\mathfrak{g}^{\mathbf{d}}\label{n9a}%
\end{equation}
the nice formula promised and that clearly demonstrates that the objects
$\mathbf{t}_{\mathbf{da}}=\eta_{\mathbf{ac}}\eta_{\mathbf{dl}}\mathbf{t}%
^{\mathbf{c}}\underset{g}{\lrcorner}\mathfrak{g}^{\mathbf{l}}$ are components
of a legitimate gravitational energy-momentum tensor tensor field
$\mathbf{t}=\mathbf{t}_{\mathbf{da}}\mathfrak{g}^{\mathbf{d}}\otimes
\mathfrak{g}^{\mathbf{a}}\in\sec T_{0}^{2}M$. We observe moreover that
\begin{equation}
\mathbf{t}^{\mathbf{da}}-\mathbf{t}^{\mathbf{ad}}=2\boldsymbol{\partial}%
\cdot\boldsymbol{\partial}\mathfrak{g}^{\mathbf{d}}\underset{g}{\lrcorner
}\mathfrak{g}^{\mathbf{a}},\label{n10}%
\end{equation}
i.e., the energy-momentum tensor of the gravitational field in not symmetric.
As shown in \cite{fr} this is important in order to have a total angular
momentum conservation law for the system consisting of the gravitational plus
the matter fields. At least observe that $\mathbf{t}^{\mathbf{d}}%
=\mathfrak{t}^{\mathbf{d}}$ when the potentials are choosen in the Lorenz gauge.

\section{Hamilton Formalism}

If we define as usual the canonical momenta associated to the potentials
$\{\mathfrak{g}^{a}\}$ by $\mathfrak{p}_{\mathbf{a}}=\partial\mathcal{L}%
_{g}/\partial d\mathfrak{g}^{\mathbf{a}}=\underset{g}{\star}\mathcal{S}%
_{\mathbf{a}}$ and suppose that this equation can be solved for the
$d\mathfrak{g}^{\mathbf{a}}$ as function of the $\mathfrak{p}_{\mathbf{a}}$ we
can introduce a Legendre transformation with respect to the \ fields
\ $d\mathfrak{g}^{\mathbf{a}}$ by%

\begin{equation}
\mathbf{L}:(\mathfrak{g}^{\mathbf{\alpha}},\mathfrak{p}_{\mathbf{\alpha}%
})\mapsto\mathbf{L}(\mathfrak{g}^{\mathbf{\alpha}},\mathfrak{p}%
_{\mathbf{\alpha}})=d\mathfrak{g}^{\mathbf{\alpha}}\wedge p_{\mathbf{\alpha}%
}-\mathcal{L}_{g}(\mathfrak{g}^{\mathbf{\alpha}},d\mathfrak{g}^{\mathbf{\alpha
}}(\mathfrak{p}_{\mathbf{\alpha}})) \label{h3}%
\end{equation}

We write in what follows $\mathfrak{L}_{g}(\mathfrak{g}^{\mathbf{\alpha}%
},\mathfrak{p}_{\alpha}):=\mathcal{L}_{g}(\mathfrak{g}^{\mathbf{\alpha}%
},d\mathfrak{g}^{\mathbf{\alpha}}(\mathfrak{p}_{\alpha}))$ and observe that
defining\footnote{We use only constrained variations of the $\mathfrak{g}%
^{\mathbf{a}}$, which as already recalled in Section 2 do not change the
metric field $\boldsymbol{g}$.}
\begin{equation}
\frac{%
%TCIMACRO{\TeXButton{delta}{\mbox{\boldmath{$\delta$}}}}%
%BeginExpansion
\mbox{\boldmath{$\delta$}}%
%EndExpansion
\mathfrak{L}_{g}(\mathfrak{g}^{\mathbf{\alpha}},\mathfrak{p}_{\mathbf{\alpha}%
})}{%
%TCIMACRO{\TeXButton{delta}{\mbox{\boldmath{$\delta$}}}}%
%BeginExpansion
\mbox{\boldmath{$\delta$}}%
%EndExpansion
\mathfrak{g}^{\mathbf{\alpha}}}:=-d\mathfrak{p}_{\mathbf{\alpha}}%
-\frac{\partial\mathbf{L}}{\partial\mathfrak{g}^{\mathbf{\alpha}}}%
,\text{....}\frac{%
%TCIMACRO{\TeXButton{delta}{\mbox{\boldmath{$\delta$}}}}%
%BeginExpansion
\mbox{\boldmath{$\delta$}}%
%EndExpansion
\mathfrak{L}_{g}(\mathfrak{g}^{\mathbf{\alpha}},\mathfrak{p}_{\mathbf{\alpha}%
})}{%
%TCIMACRO{\TeXButton{delta}{\mbox{\boldmath{$\delta$}}}}%
%BeginExpansion
\mbox{\boldmath{$\delta$}}%
%EndExpansion
\mathfrak{p}_{\mathbf{\alpha}}}:=d\mathfrak{g}^{\mathbf{\alpha}}%
-\frac{\partial\mathbf{L}}{\partial\mathfrak{p}_{\mathbf{\alpha}}}\nonumber
\end{equation}
we can obtain (see details in \cite{fr})
\begin{equation}%
%TCIMACRO{\TeXButton{delta}{\mbox{\boldmath{$\delta$}}}}%
%BeginExpansion
\mbox{\boldmath{$\delta$}}%
%EndExpansion
\mathfrak{g}^{\mathbf{\alpha}}\wedge\frac{%
%TCIMACRO{\TeXButton{delta}{\mbox{\boldmath{$\delta$}}}}%
%BeginExpansion
\mbox{\boldmath{$\delta$}}%
%EndExpansion
\mathcal{L}_{g}(\mathfrak{g}^{\mathbf{\alpha}},d\mathfrak{g}^{\mathbf{\alpha}%
})}{%
%TCIMACRO{\TeXButton{delta}{\mbox{\boldmath{$\delta$}}}}%
%BeginExpansion
\mbox{\boldmath{$\delta$}}%
%EndExpansion
\mathfrak{g}^{\mathbf{\alpha}}}=%
%TCIMACRO{\TeXButton{delta}{\mbox{\boldmath{$\delta$}}}}%
%BeginExpansion
\mbox{\boldmath{$\delta$}}%
%EndExpansion
\mathfrak{g}^{\mathbf{\alpha}}\wedge\left(  \frac{%
%TCIMACRO{\TeXButton{delta}{\mbox{\boldmath{$\delta$}}}}%
%BeginExpansion
\mbox{\boldmath{$\delta$}}%
%EndExpansion
\mathfrak{L}_{g}(\mathfrak{g}^{\mathbf{\alpha}},\mathfrak{p}_{\mathbf{\alpha}%
})}{%
%TCIMACRO{\TeXButton{delta}{\mbox{\boldmath{$\delta$}}}}%
%BeginExpansion
\mbox{\boldmath{$\delta$}}%
%EndExpansion
\mathfrak{g}^{\mathbf{\alpha}}}\right)  +\left(  \frac{%
%TCIMACRO{\TeXButton{delta}{\mbox{\boldmath{$\delta$}}}}%
%BeginExpansion
\mbox{\boldmath{$\delta$}}%
%EndExpansion
\mathfrak{L}_{g}(\mathfrak{g}^{\mathbf{\alpha}},\mathfrak{p}_{\mathbf{\alpha}%
})}{%
%TCIMACRO{\TeXButton{delta}{\mbox{\boldmath{$\delta$}}}}%
%BeginExpansion
\mbox{\boldmath{$\delta$}}%
%EndExpansion
\mathfrak{p}_{\mathbf{\alpha}}}\right)  \wedge%
%TCIMACRO{\TeXButton{delta}{\mbox{\boldmath{$\delta$}}}}%
%BeginExpansion
\mbox{\boldmath{$\delta$}}%
%EndExpansion
\mathfrak{p}_{\mathbf{\alpha}}. \label{h6'}%
\end{equation}

To define the Hamiltonian form, we need something to act the role of time for
our manifold, and we choose this\ `time' to be given by the flow of an
arbitrary timelike vector field $\mathbf{Z}\in\sec TM$ such that $%
%TCIMACRO{\TeXButton{g}{\slg}}%
%BeginExpansion
\slg
%EndExpansion
(\mathbf{Z,Z)}=1$. Moreover, we define $Z=%
%TCIMACRO{\TeXButton{g}{\slg}}%
%BeginExpansion
\slg
%EndExpansion
(\mathbf{Z,)}\in\sec\bigwedge^{1}T^{\ast}M\hookrightarrow\mathcal{C\ell
(}g,M\mathcal{)}$. With this choice, the variation $%
%TCIMACRO{\TeXButton{delta}{\mbox{\boldmath{$\delta$}}}}%
%BeginExpansion
\mbox{\boldmath{$\delta$}}%
%EndExpansion
$ is generated by the Lie derivative $\pounds _{\mathbf{Z}}$. Using Cartan's
`magical formula', we have
\begin{equation}%
%TCIMACRO{\TeXButton{delta}{\mbox{\boldmath{$\delta$}}}}%
%BeginExpansion
\mbox{\boldmath{$\delta$}}%
%EndExpansion
\mathfrak{L}_{g}=\pounds _{\mathbf{Z}}\mathfrak{L}_{g}=d(Z\lrcorner
\mathfrak{L}_{g})+Z\lrcorner d\mathfrak{L}_{g}=d(Z\lrcorner\mathfrak{L}_{g}).
\label{h7}%
\end{equation}
and after some algebra we get
\begin{equation}
d(Z\lrcorner\mathfrak{L}_{g})=d(\pounds _{\mathbf{Z}}\mathfrak{g}%
^{\mathbf{\alpha}}\wedge\mathfrak{p}_{\mathbf{\alpha}})+\pounds _{\mathbf{Z}%
}\mathfrak{g}^{\mathbf{\alpha}}\wedge\frac{%
%TCIMACRO{\TeXButton{delta}{\mbox{\boldmath{$\delta$}}}}%
%BeginExpansion
\mbox{\boldmath{$\delta$}}%
%EndExpansion
\mathfrak{L}_{g}}{%
%TCIMACRO{\TeXButton{delta}{\mbox{\boldmath{$\delta$}}}}%
%BeginExpansion
\mbox{\boldmath{$\delta$}}%
%EndExpansion
\mathfrak{g}^{\mathbf{\alpha}}}+\pounds _{\mathbf{Z}}\mathfrak{p}%
_{\mathbf{\alpha}}\wedge\left(  \frac{%
%TCIMACRO{\TeXButton{delta}{\mbox{\boldmath{$\delta$}}}}%
%BeginExpansion
\mbox{\boldmath{$\delta$}}%
%EndExpansion
\mathfrak{L}}{%
%TCIMACRO{\TeXButton{delta}{\mbox{\boldmath{$\delta$}}}}%
%BeginExpansion
\mbox{\boldmath{$\delta$}}%
%EndExpansion
\mathfrak{p}_{\mathbf{\alpha}}}\right)  \label{h8}%
\end{equation}
and also%
\begin{equation}
d(\pounds _{\mathbf{Z}}\mathfrak{g}^{\mathbf{\alpha}}\wedge\mathfrak{p}%
_{\mathbf{\alpha}}-Z\lrcorner\mathfrak{L}_{g})=\pounds _{\mathbf{Z}%
}\mathfrak{g}^{\mathbf{\alpha}}\wedge\frac{%
%TCIMACRO{\TeXButton{delta}{\mbox{\boldmath{$\delta$}}}}%
%BeginExpansion
\mbox{\boldmath{$\delta$}}%
%EndExpansion
\mathcal{L}_{g}}{%
%TCIMACRO{\TeXButton{delta}{\mbox{\boldmath{$\delta$}}}}%
%BeginExpansion
\mbox{\boldmath{$\delta$}}%
%EndExpansion
\mathfrak{g}^{\mathbf{\alpha}}}. \label{h9}%
\end{equation}
Now, we define the \textit{Hamiltonian} $3$-form by%
\begin{equation}
\mathcal{H(\mathfrak{g}^{\mathbf{\alpha}}},\mathcal{\mathfrak{p}%
_{\mathbf{\alpha}})}:=\pounds _{\mathbf{Z}}\mathfrak{g}^{\mathbf{\alpha}%
}\wedge\mathfrak{p}_{\mathbf{\alpha}}-Z\lrcorner\mathfrak{L}_{g}. \label{h10}%
\end{equation}
We immediately have taking into account Eq.(\ref{h9}) that, when the field
equations for the \textit{free }gravitational field are satisfied (i.e., when
the Euler-Lagrange functional is null, $%
%TCIMACRO{\TeXButton{delta}{\mbox{\boldmath{$\delta$}}}}%
%BeginExpansion
\mbox{\boldmath{$\delta$}}%
%EndExpansion
\mathcal{L}_{g}/%
%TCIMACRO{\TeXButton{delta}{\mbox{\boldmath{$\delta$}}}}%
%BeginExpansion
\mbox{\boldmath{$\delta$}}%
%EndExpansion
\mathfrak{g}^{\mathbf{\alpha}}=0$) that
\begin{equation}
d\mathcal{H}=0. \label{h11}%
\end{equation}
Thus $\mathcal{H}$ is a \textit{conserved} Noether current. We next write
\begin{equation}
\mathcal{H}=Z^{\mathbf{\alpha}}\mathcal{H}_{\mathbf{\alpha}}+dB \label{h12}%
\end{equation}
We can show (details in \cite{fr}) that $\mathcal{H}_{\mathbf{\alpha}}=-%
%TCIMACRO{\TeXButton{delta}{\mbox{\boldmath{$\delta$}}}}%
%BeginExpansion
\mbox{\boldmath{$\delta$}}%
%EndExpansion
\mathcal{L}_{g}/%
%TCIMACRO{\TeXButton{delta}{\mbox{\boldmath{$\delta$}}}}%
%BeginExpansion
\mbox{\boldmath{$\delta$}}%
%EndExpansion
\mathfrak{g}^{\mathbf{\alpha}}$ and $B=Z^{\mathbf{a}}\mathfrak{p}_{\mathbf{a}%
}$ and now we investigate the meaning of the boundary term\footnote{More
details on possible choices of the boundary term for different physical
situations may be found in \cite{meng}.} $B$. Consider an arbitrary spacelike
hypersuface $\sigma$. Then, we define%
\begin{equation}
\mathbf{H}=%
%TCIMACRO{\dint \nolimits_{\sigma}}%
%BeginExpansion
{\displaystyle\int\nolimits_{\sigma}}
%EndExpansion
(Z^{\alpha}\mathcal{H}_{\alpha}+dB)=%
%TCIMACRO{\dint \nolimits_{\sigma}}%
%BeginExpansion
{\displaystyle\int\nolimits_{\sigma}}
%EndExpansion
Z^{\alpha}\mathcal{H}_{\alpha}+%
%TCIMACRO{\dint \nolimits_{\partial\sigma}}%
%BeginExpansion
{\displaystyle\int\nolimits_{\partial\sigma}}
%EndExpansion
B.\nonumber
\end{equation}
If we recall that $\mathcal{H}_{\mathbf{\alpha}}=-%
%TCIMACRO{\TeXButton{delta}{\mbox{\boldmath{$\delta$}}}}%
%BeginExpansion
\mbox{\boldmath{$\delta$}}%
%EndExpansion
\mathcal{L}_{g}/%
%TCIMACRO{\TeXButton{delta}{\mbox{\boldmath{$\delta$}}}}%
%BeginExpansion
\mbox{\boldmath{$\delta$}}%
%EndExpansion
\mathfrak{g}^{\mathbf{\alpha}}$ we see that the first term in the above
equation is null when the field equations (for the free gravitational field)
are satisfied and we are thus left with
\begin{equation}
\mathbf{E}=%
%TCIMACRO{\dint \nolimits_{\partial\sigma}}%
%BeginExpansion
{\displaystyle\int\nolimits_{\partial\sigma}}
%EndExpansion
B, \label{h26}%
\end{equation}
which is called the quasi local energy \cite{sza}.

Now, if $\{%
%TCIMACRO{\TeXButton{e}{\sle}}%
%BeginExpansion
\sle
%EndExpansion
_{\mathbf{\alpha}}\}$ is the dual basis of $\{\mathfrak{g}^{\mathbf{\alpha}%
}\}$ we have $\mathfrak{g}^{\mathbf{0}}(%
%TCIMACRO{\TeXButton{e}{\sle}}%
%BeginExpansion
\sle
%EndExpansion
_{\mathbf{i}})=0,$ $\mathbf{i=}1,2,3$ and if we take $\mathbf{Z=}$ $%
%TCIMACRO{\TeXButton{e}{\sle}}%
%BeginExpansion
\sle
%EndExpansion
_{\mathbf{0}}$ orthogonal to the hypersurface $\sigma$, such that for each
$p\in\sigma$, $T\sigma_{p}$ is generated by $\{%
%TCIMACRO{\TeXButton{e}{\sle}}%
%BeginExpansion
\sle
%EndExpansion
_{\mathbf{i}}\}$ and we get recalling that $\mathfrak{p}_{\mathbf{\alpha}%
}=\underset{%
%TCIMACRO{\TeXButton{sig}{\sitg}}%
%BeginExpansion
\sitg
%EndExpansion
}{\star}\mathcal{S}_{\mathbf{\alpha}}$ that
\begin{equation}
\mathbf{E}=%
%TCIMACRO{\dint \nolimits_{\partial\sigma}}%
%BeginExpansion
{\displaystyle\int\nolimits_{\partial\sigma}}
%EndExpansion
\underset{%
%TCIMACRO{\TeXButton{sig}{\sitg}}%
%BeginExpansion
\sitg
%EndExpansion
}{\star}\mathcal{S}_{0}, \label{h128}%
\end{equation}
which we recognize as being the same conserved quantity as the one defined by
Eq.(\ref{energy}).

The relation of the energy defined by Eq.(\ref{h128}) with the energy concept
defined in \textit{ADM} formalism \cite{adm} can be seen as follows
\cite{wallner}. Instead of choosing an arbitrary unit timelike vector field
$\mathbf{Z}$, start with a global timelike vector field $\mathbf{n}\in\sec TM$
such that \ $n=%
%TCIMACRO{\TeXButton{slg}{\slg}}%
%BeginExpansion
\slg
%EndExpansion
(\mathbf{n,\,})=N^{2}dt\in\sec\bigwedge^{1}T^{\ast}M\hookrightarrow
\mathcal{C\ell(}M,g)$, with $N:\mathbb{R\supset I\rightarrow R}$, a positive
function called the lapse function of $M$. Then $n\wedge dn=0$ and according
to Frobenius theorem, $n$ induces a foliation of $M$, i.e., topologically it
is $M=\mathbb{I\times}\sigma_{t},$ where $\sigma_{t}$ is a spacelike
hypersurface with normal given by $\mathbf{n}$. Now, we can decompose any
$A\in\sec\bigwedge^{p}T^{\ast}M\hookrightarrow\mathcal{C\ell(}M,g)$ into\ a
tangent component $\underline{A}$ to $\sigma_{t}$ and an orthogonal component
$\ ^{\perp}A$ to $\sigma_{t}$ by
\begin{equation}
A=\underline{A}+\text{ }^{\perp}A\text{,} \label{h43}%
\end{equation}
where%
\begin{equation}
\underline{A}:=n\lrcorner(dt\wedge A)\text{, }^{\perp}A=dt\wedge A_{\perp
}\text{,.. }A_{\perp}\text{ }\text{:}=n\lrcorner A. \label{h44}%
\end{equation}
Introduce also the parallel component \underline{$d$} of the differential
operator $d$ by:%
\begin{equation}
\underline{d}A:=n\lrcorner(dt\wedge dA) \label{h45}%
\end{equation}
from where it follows (taking into account Cartan's magical formula) that
\begin{equation}
dA=dt\wedge(\pounds _{\mathbf{n}}\underline{A}-\underline{d}A_{\perp
})+\underline{dA}. \label{h46}%
\end{equation}
Call \
\begin{equation}%
%TCIMACRO{\TeXButton{m}{\slm}}%
%BeginExpansion
\slm
%EndExpansion
:=-%
%TCIMACRO{\TeXButton{g}{\slg}}%
%BeginExpansion
\slg
%EndExpansion
+\boldsymbol{n}\otimes\boldsymbol{n}=\underline{\mathfrak{g}}^{i}%
\otimes\underline{\mathfrak{g}}_{i},\nonumber
\end{equation}
(where $\boldsymbol{n=}n/N$) the first fundamental form on $\sigma_{t}$ and
next introduce the Hodge dual operator associated to $%
%TCIMACRO{\TeXButton{m}{\slm}}%
%BeginExpansion
\slm
%EndExpansion
$, acting on the (horizontal forms) forms $\underline{A}$ by
\begin{equation}
\underset{%
%TCIMACRO{\TeXButton{sm}{\sslm}}%
%BeginExpansion
\sslm
%EndExpansion
}{\star}\underline{A}:=\underset{%
%TCIMACRO{\TeXButton{sg}{\sslg}}%
%BeginExpansion
\sslg
%EndExpansion
}{\star}(\frac{n}{N}\wedge\underline{A}). \label{H47}%
\end{equation}
At this point, we come back to the Lagrangian density Eq.(\ref{h10}) and,
proceeding like above, but now leaving $%
%TCIMACRO{\TeXButton{delta}{\mbox{\boldmath{$\delta$}}}}%
%BeginExpansion
\mbox{\boldmath{$\delta$}}%
%EndExpansion
n^{\mathbf{\alpha}}$ to be non null, we eventually arrive at the following
Hamiltonian density
\begin{equation}
\mathcal{H(\underline{\mathfrak{g}}}^{i},\underline{\mathfrak{p}}%
_{i}\mathcal{)}=\pounds _{\mathbf{n}}\underline{\mathfrak{g}}^{i}%
\wedge\underset{%
%TCIMACRO{\TeXButton{sm}{\sslm}}%
%BeginExpansion
\sslm
%EndExpansion
}{\star}\underline{\mathfrak{p}}_{i}-\mathcal{K}_{g}, \label{h50bis}%
\end{equation}
where%
\begin{equation}
\mathfrak{g}^{i}-\underline{\mathfrak{g}}^{i}=dt\wedge(n\lrcorner
\mathfrak{g}^{i})=n^{i}dt, \label{h52}%
\end{equation}
and where $\mathcal{K}_{g}$ depends on $(n,\underline{d}%
n,\underline{\mathfrak{g}}^{i},\underline{d\mathfrak{g}}^{i}%
,\pounds _{\mathbf{n}}\underline{\mathfrak{g}}^{i})$. We can show (after some
tedious but straightforward algebra that $\mathcal{H(\underline{\mathfrak{g}}%
}^{i},\underline{\mathfrak{p}}_{i}\mathcal{)}$ can be put into the form%
\begin{equation}
\mathcal{H}=n^{i}\mathcal{H}_{i}+\underline{d}B^{\prime}, \label{h51}%
\end{equation}
with as before $\mathcal{H}_{i}=-%
%TCIMACRO{\TeXButton{delta}{\mbox{\boldmath{$\delta$}}}}%
%BeginExpansion
\mbox{\boldmath{$\delta$}}%
%EndExpansion
\mathcal{L}_{g}/%
%TCIMACRO{\TeXButton{delta}{\mbox{\boldmath{$\delta$}}}}%
%BeginExpansion
\mbox{\boldmath{$\delta$}}%
%EndExpansion
\mathfrak{g}^{i}=-%
%TCIMACRO{\TeXButton{delta}{\mbox{\boldmath{$\delta$}}}}%
%BeginExpansion
\mbox{\boldmath{$\delta$}}%
%EndExpansion
\mathcal{K}_{g}/%
%TCIMACRO{\TeXButton{delta}{\mbox{\boldmath{$\delta$}}}}%
%BeginExpansion
\mbox{\boldmath{$\delta$}}%
%EndExpansion
n^{i}$ and
\begin{equation}
B^{\prime}=-N\underline{\mathfrak{g}}_{i}\wedge\underset{%
%TCIMACRO{\TeXButton{sm}{\sslm}}%
%BeginExpansion
\sslm
%EndExpansion
}{\star}\underline{d}\underline{\mathfrak{g}}^{i} \label{h55}%
\end{equation}
Then, on shell, i.e., when the field equations are satisfied we get
\begin{equation}
\mathbf{E}^{\prime}\mathbf{=-}%
%TCIMACRO{\dint \nolimits_{\partial\sigma_{t}}}%
%BeginExpansion
{\displaystyle\int\nolimits_{\partial\sigma_{t}}}
%EndExpansion
N\underline{\mathfrak{g}}_{i}\wedge\underset{%
%TCIMACRO{\TeXButton{sm}{\sslm}}%
%BeginExpansion
\sslm
%EndExpansion
}{\star}\underline{d}\underline{\mathfrak{g}}^{i} \label{h56}%
\end{equation}
which is exactly the \textit{ADM} energy, as can be seen if we take into
account that taking $\partial\sigma_{t}$ as a twosphere at infinity, we have
(using coordinates in the ELP gauge) $\underline{\mathfrak{g}_{i}}%
=h_{ij}\underline{d}\mathrm{x}^{j}$ and $h_{ij}$, $N\rightarrow1$. Then%
\begin{equation}
\underline{\mathfrak{g}}_{i}\wedge\underset{%
%TCIMACRO{\TeXButton{sm}{\sslm}}%
%BeginExpansion
\sslm
%EndExpansion
}{\star}\underline{d}\underline{\mathfrak{g}}^{i}=h^{ij}(\frac{\partial
h_{ij}}{\partial\mathrm{x}^{k}}-\frac{\partial h_{ik}}{\partial\mathrm{x}^{j}%
})\underset{%
%TCIMACRO{\TeXButton{sm}{\sslm}}%
%BeginExpansion
\sslm
%EndExpansion
}{\star}\underline{\mathfrak{g}}^{k} \label{h57}%
\end{equation}
and under the above conditions we have the \textit{ADM} formula
\begin{equation}
\mathbf{E}^{\prime}=%
%TCIMACRO{\dint \nolimits_{\partial\sigma_{t}}}%
%BeginExpansion
{\displaystyle\int\nolimits_{\partial\sigma_{t}}}
%EndExpansion
\left(  \frac{\partial h_{ik}}{\partial\mathrm{x}^{i}}-\frac{\partial h_{ik}%
}{\partial\mathrm{x}^{k}}\right)  \underset{%
%TCIMACRO{\TeXButton{sm}{\sslm}}%
%BeginExpansion
\sslm
%EndExpansion
}{\star}\underline{\mathfrak{g}}^{k}. \label{h58}%
\end{equation}
which, as is well known , is positive definite\footnote{See a nice proof in
\cite{wallner}.}. If we choose $n=\mathfrak{g}^{0}$ it may happen that
$\mathfrak{g}^{0}\wedge d\mathfrak{g}^{0}\neq0$ \ and thus it does not
determine a spacelike hypersurface $\sigma_{t}$. However all algebraic
calculations above up to Eq.(\ref{h55}) are valid (and of course,
$\mathfrak{g}^{k}=\underline{\mathfrak{g}}^{k}$). So, if we take a spacelike
hypersurface $\sigma$ such that at spatial infinity the $\mathfrak{e}_{i}$
($\mathfrak{g}^{k}(\mathfrak{e}_{i})=\delta_{i}^{k}$) are tangent to $\sigma$,
and $\mathfrak{e}_{0}\rightarrow\partial/\partial t$ is orthogonal to$\sigma$,
then we have $\mathbf{E=E}^{\prime}$ since in this case
$-N\underline{\mathfrak{g}}_{i}\wedge\underset{%
%TCIMACRO{\TeXButton{sm}{\sslm}}%
%BeginExpansion
\sslm
%EndExpansion
}{\star}\underline{d}\underline{\mathfrak{g}}^{i}\rightarrow-\mathfrak{g}%
_{i}\wedge\underset{%
%TCIMACRO{\TeXButton{sm}{\sslm}}%
%BeginExpansion
\sslm
%EndExpansion
}{\star}(\mathfrak{g}^{0}\wedge\underset{%
%TCIMACRO{\TeXButton{sm}{\sslm}}%
%BeginExpansion
\sslm
%EndExpansion
}{\star}d\mathfrak{g}^{i})$ which as can be easily verified (see
Eq.(\ref{14})) is the asymptotic value of $\underset{%
%TCIMACRO{\TeXButton{sg}{\sslg}}%
%BeginExpansion
\sslg
%EndExpansion
}{\star}\mathcal{S}^{0}$ (taking into account that at spatial infinity
$d\mathfrak{g}^{0}\rightarrow0$)

\section{Conclusions}

In this paper we recalled that a gravitational field generated by a given
energy-momentum distribution can be represented by distinct geometrical
structures and if we prefer, we can even dispense all those geometrical
structures and simply represent the gravitational field as a field in the
Faraday's sense living in Minkowski spacetime. The explicit Lagrangian density
for this theory has been given in a Maxwell like form and shown to be
equivalent to Einstein's equations in a precise mathematical sense. We
identify a legitimate energy-momentum tensor for the gravitational field which
can be expressed through a really nice formula, namely Eq.(\ref{n9}). We hope
that our study clarifies the real difference between mathematical models and
physical reality and leads people to think about the real physical nature of
the gravitational field (and also of the electromagnetic field\footnote{As
suggested, e.g., by the works of Laughlin \cite{laughlin} and Volikov
\cite{volovik}. Of course,, it may be necessary to explore also other ideas,
like e.g., existence of branes in string theory. But this is a subject for
another publication.}). We discuss also an Hamiltonian formalism for our
theory and the concept of energy defined by Eq.(\ref{energy}) and the one
given by the ADM formalism, which are shown to coincide

\end{document}